\documentclass[acus]{JAC2000}

\usepackage{graphicx}
\newcommand{\ben}{\begin{equation}}
\newcommand{\een}{\end{equation}}

\begin{document}
\title{2D SIMULATION OF HIGH-EFFICIENCY CROSS-FIELD RF POWER SOURCES\thanks{This work
was supported by the U.S. Department of Energy contract
DE-AC03-76SF00515.}}

\author{Valery A. Dolgashev, Sami G. Tantawi\thanks{Also with the
Communications and Electronics Department, Cairo University,
Giza, Egypt.}, SLAC, Stanford, CA 94309, USA}

\maketitle

\section{Introduction}
In a cross field device\cite{magnetron} such as magnetron or
cross field amplifier electrons move in crossed magnetic and
electric fields. Due to synchronism between electron drift
velocity and phase velocity of RF wave, the wave bunches the
beam, electron spokes are formed and the bunched electrons are
decelerated by the RF field. Such devices have high efficiency
(up to 90\%), high output power and relatively low cost.
Electrical design of the cross-field devices is difficult. The
problem is 2D (or 3D) and highly nonlinear. It has complex
geometry and strong space charge effects. Recently, increased
performance of computers and availability of Particle-In-Cell
(PIC) codes\cite{magic,Eppley}, have made possible the design of
relatively low efficiency devices such as relativistic magnetrons
or cross field amplifiers \cite{Chen}. Simulation of high
efficiency ($\sim 90\%$) devices is difficult due to the long
transient process of starting oscillations. Use of PIC codes for
design of such devices is not practical. In this report we
describe a frequency domain method that developed for simulating
high efficiency cross-field devices. In the method, we consider
steady-state interaction of particles with the modes of RF cavity
at dominant frequency. Self-consistency of the solution is reached
by iterations until power balance is achieved.
\section{Physical model}

Cross-field devices consist of a cathode and a surrounding anode.
The structure is a cavity with a set of resonant eigenmodes.
Macroparticles are emitted from the cathode and moved by forces of
electromagnetic fields. The electromagnetic fields are determined
by applied external electric potential between anode and cathode,
oscillating field of cavity modes, and space charge fields. We
use geometry with arbitrary piece-wise planar boundaries. In order
to solve the electrostatic and electrodynamic problems, we apply
methods  that do not require mesh generation. Interaction with
magnetic field is determined by uniform magnetic field $H_z$
which is  parallel to $z$-axis and orthogonal to the plane of
simulation. There are several assumptions that we use to simplify
the problem. These assumptions are based on the working regime of
the devices that we want to simulate. Devices will have low
current density, are non-relativistic, and have resonant systems
with a relatively low density of the cavity modes.  Hence, we can
neglect magnetic fields due to space charge and cavity modes. We
can also use cavity modes with eigen-frequencies close to the
working frequency. \vspace{-1mm}
\subsection{Basic equations}
\vspace{-2mm}
 We are solving a steady state problem of
electron beam flow in self-consistent electromagnetic fields.
Total fields are superposition of static electric $\vec{E}'$ and
magnetic $\vec{H}'$ fields, and  ``oscillating'' electric
$\vec{E}(\omega)$ and magnetic $\vec{H}(\omega)$ fields as
\[ \vec{E}(t) = \vec{E'} +
\Re e\{{\vec{E}(\omega)e^{j\omega t}}\},~ \vec{H}(t) = \vec{H'} +
\Re e \{{\vec{H}(\omega)e^{j\omega t}}\}. \]
 Here $\omega$ is angular frequency, $t$ is time.
We separate the electrodynamic problem into two parts. The first
part -- electrostatic potential $\Phi$ is generated by
``external'' anode-cathode potential and by the static component
of  the space-charge electric fields. The second part -- the
dynamic electromagnetic fields have a harmonic $e^{j\omega t}$
time ($t$) dependence.
\vspace{-1mm}
\subsection{Static fields }
\vspace{-2mm}
We find the static electric field from $\vec{E'}= - \nabla \Phi$,
using the {\it Poisson equation}: \vspace{-3mm}
\begin{equation}
\nabla^2 \Phi = -\frac{\rho}{\epsilon_0}, \label{phi_rho}
\end{equation}
\vspace{-1mm} where $\nabla$ is the gradient operator, $\rho$ is
volume charge density averaged over  oscillation period
$T=2\pi/\omega$. $\epsilon_0$ is the electric permittivity of the
vacuum.
%
\vspace{-1mm}
\subsection{Oscillating fields}
\vspace{-2mm}
 To solve the second part of the problem,
we write the time harmonic {\it Maxwell equations} as
 \ben
\nabla \times \vec{E} = -j \omega \mu_0 \vec{H}, ~ \nabla \times
\vec{H} = j \omega \epsilon_0 \vec{E} + \vec{J_\omega}. \een
Here $\vec{J}_\omega$ is electric current density,  $\mu_0$ is the
magnetic permeability of  vacuum. Oscillating fields inside a
cavity are expanded in terms of the cavity
 eigenmodes ($\vec{E_s}$, $\vec{H_s})$ and the {\it fast oscillating}
 electric potential $\varphi_\omega$ as
\vspace{-1mm} \ben \vec{E} = \sum_s A_s \vec{E_s} - \nabla
\varphi_\omega,~ \vec{H} = \sum_s B_s \vec{H_s}.
\label{ehExpansion} \vspace{-2mm} \een

Here $s$ is mode index, $A_s$ and $B_s$ are the
eigenmode amplitudes. Using the expansion (\ref{ehExpansion}) we
get the {\it Poisson equation} for the potential: \vspace{-3mm}
\ben \nabla^2 \varphi = \frac{\nabla \cdot \vec{J_\omega}}{j
\omega \epsilon_0} = - \frac{\rho_\omega}{\epsilon_0}, \een
where $\rho_\omega$ is the oscillating space-charge density.
Amplitudes of the electric field expansion are given by
\ben A_s =\frac{ \omega}{j(\omega^2 - \omega_s^2)}  \frac{\int_V
\vec{J_\omega}\vec{E}_s^* dV}{\epsilon_0 \int_V
\vec{E}_s\vec{E}_s^* dV}. \label{A_s} \een
Here $\omega_s$ is the mode eigen-frequency of the mode , $V$ is
the cavity volume.
\vspace{-1mm}
\subsection{Equation of motion}
\vspace{-2mm}
Equation of motion for an electron in crossed-fields is \ben
\frac{d\vec{p}}{dt} = q_e { \vec{E}(t) + \mu_0 \vec{v}\ \times
{H_z}}, \label{eq_motion} \een
where $\vec{p}$ is the relativistic momentum, $q_e$ is the
charge, and $\vec{v}$ is the velocity of the electron. Current
density induced by the electron motion is $\vec{J} = q_e \vec{v}
\delta(\vec{r}),$ where $\vec{r}$ is the position vector of the
electron, and $\delta$ is the {\it Kronecker delta function}.
\section{Numerical methods}
We created several separate program  modules to simulate a
cross-field device. First is an { \it RF field solver} that
calculates eigenmodes  and eigen-frequencies in the cavity;
second is the {\it Poisson solver} that finds electric fields due
to external potential, static space charge, and oscillating space
charge; and third, the {\it tracking module} that performs
tracking of electrons through electromagnetic fields. For
simulation, we consider an arbitrary, piecewise bounded 2D
geometry. \vspace{-1mm} \vspace{-3mm}
\subsection{Planar geometry}
\vspace{-4mm}
\begin{figure}[htb]
\centering
\includegraphics*[width=50mm]{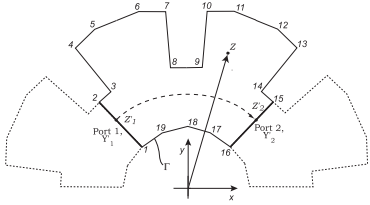}
\caption{\small Planar geometry.} \label{pl_pssn}
\end{figure}
The geometry is cylindrical (uniform in the $z$-direction) as
illustrated on Fig. \ref{pl_pssn}. It consists of planar
sidewalls and apertures. The geometry in the  $x,y$ plane can be
described by a set of points $z_s = (x_{s}, y_{s})$, where $s =
1,2...,N';$ here $N'$ is the total number of sidewalls and
apertures. Periodic boundary conditions are applied to the
apertures. The periodic boundary allows us to use only part of the
structure and significantly reduce simulation time.  In the
particular case shown in Fig. \ref{pl_pssn} the geometry has
$N'=19$ sidewalls, two apertures (ports) with starting points $p
= 1,~15$, and the cathode and anode determined by
$s=16,17,18,19,1$ and $s=2,3,...,15$ respectively. \vspace{-1mm}
\subsection{RF field solver}
\vspace{-2mm}
 The description of the RF solver that is used in
this method is published in \cite{epac2000}. Here we briefly
outline its properties. We use the scattering matrix approach
\cite{icap98} to calculate the dispersion parameters of the
periodic 2D structure, it's  resonant frequencies, and the
corresponding fields. The fields are described by functional
expansion. Boundary contour mode-matching is applied in a
piecewise bounded 2D region is applied to obtain the scattering
matrix and field amplitudes \cite{bessel}. The Galerkin method is
used for the mode-matching procedure. The geometry is divided
into regions, and electromagnetic fields in each region are
expanded in series of plane waves or (for low frequencies){ \it
Bessel functions}. Scattering matrices from the regions are
combined using the generalized scattering matrix technique.
Resonant and periodic boundary conditions \cite{icap98} are used
to obtain resonant frequencies, dispersion parameters, and
corresponding fields. We calculate the electric fields on a polar
grid (only in the region of field-particle interaction), in order
to speed up calculation of fields for the macroparticle tracking.
To obtain field at the macroparticle position we use 2D spline
interpolation.
\subsection{Poisson solver}
\vspace{-2mm}
\label{PoissonSolver} We use an efficient method for solving the
Poisson equation for electric fields in a 2-D, arbitrarily shaped
geometry. The solution is based on the  method of moments.
Point-matching in a piecewise bounded 2D region is applied to
obtain the charge density on the boundary. The boundary's charge
density determines the fields and potentials throughout the
interior region. We use a complex representation of the fields
and potentials in the solution \cite{complex}. We apply periodic
boundary conditions to simulate the fields in the periodic
structure.
\subsubsection{Formulation} We solve equation (\ref{phi_rho}) in
2D. In the 2D case it is advantageous to represent the position
and field vector's $(x,y)$ components by a single complex
representation. We will work with functions of a complex variable
$z = x + j y$. The field strength $\widetilde{E}$ can be written
in terms of the scalar potential $\Phi = \Phi(z)$ as \vspace{-1mm}
\begin{equation}
\widetilde{E}(z) = - \frac {d \Phi^*}{ d z}. \label{e_z}
\vspace{-1mm}
\end{equation}
 Here $*$ represents the complex conjugate.
 An effective line charge $q$ (point charge in 2D geometry) has the complex potential
$\Phi =  (q /\epsilon_0) \log z $.
 We approximate the charge distribution on the boundary of
the region as a sum of ``step'' functions. We divide each
element(sidewall and aperture) of the boundary into $N_b$ straight
pieces or ``charged lines'' with uniform charge density $\sigma$
along the piece.
%
A {\it uniformly charged straight wall} with beginning and end
coordinates $z_1$ and $z_2$, respectively, will produce  a complex
potential at the point $z_w$
\begin{equation}
\Phi(z_w) =  \int_L \frac{\sigma}{\epsilon_0} \log (z-z_w) dz,
\label{rib_pot}
\end{equation}
where $L$ is the contour along the line. Equation (\ref{rib_pot})
is integrated analytically.
%
\subsubsection{Field strength of the \textsl{charged wall}} We obtain the
electric field of the {\it charged line} by substituting
 (\ref{rib_pot})
 into (\ref{e_z}):
\begin{equation}
\frac{\widetilde{E} (z_w) \epsilon_0}{\sigma} = \left[ \frac{|z_1
- z_2|}{z_1 - z_2} \log \left(\frac{z_w-z_1}{z_w-z_2}\right)
\right]^*. \label{e_z_rb}
\end{equation}
The value of the function is undefined on the line's contour.
However, for us, the fields inside the region are of interest.
Therefore, the direction of the field (for positive charge) on the
line's contour is chosen to be directed inward. Also singularities
at points $z_1$ and $z_2$ can affect the field's calculation.
Macroparticles with finite dimensions are used to avoid this
singularity.
%
\subsubsection{Periodic boundary condition}
%
We assume that
the potential and field strength are repeated on the period's
apertures (Fig. \ref{pl_pssn}). Let $z'_1 \in Y'_1$ and $z'_2 \in
Y'_2$. If we shift the region to the right so it coincides with
the next period, the coordinate $z'_1$ will be transformed into
coordinate $z'_2$. The periodic boundary condition becomes
\begin{equation}
 \Phi(z'_1)= \Phi(z'_2), ~
\frac{\partial\Phi(z'_1)}{\partial n}= -
\frac{\partial\Phi(z'_2)}{\partial n}. \label{per_e}
\end{equation}
We assume the Dirichlet condition on the sidewalls (except for
the apertures) as
\begin{equation}
\Phi(\Gamma') = \zeta (\Gamma'),~ \Gamma = \Gamma'+Y'_1 + Y'_2.
\label{dirich_per}
\end{equation}
\subsubsection{Integral equations}
For periodic boundary conditions
(\ref{per_e}) and (\ref{dirich_per}) surface charge density
$\sigma$ must satisfy the coupled integral equations
\begin{equation}
\left\{
 {\begin{array}{c}
 \int_\Gamma \log(z_w-z) \sigma(z) dz = \epsilon_0 \zeta
(z_w), ~z_w \in \Gamma',   \\
 \int_\Gamma \log(z'_1-z) \sigma(z) dz =  \int_\Gamma \log(z'_2-z) \sigma(z)
 dz, \\
\int_\Gamma \left\{ \frac {\partial \log(z'_1-z)}{\partial n_p}
 \right\} ^* \sigma(z) dz + \pi \sigma(z'_1) = \\
= -\int_\Gamma \left\{ \frac {\partial \log(z'_2-z)}{\partial
n_p}dz\right\}^* \sigma(z) dz -  \pi
 \sigma(z_2),\\
 ~z \in \Gamma, z'_1 \in Y'_1,~ z'_2 \in Y'_2,
 \label{per_e_int}
\end{array}} \right\},
\end{equation}
in which $\frac{\partial \log(z_w-z)}{\partial n_p}$ denotes the
normal derivative of  $\log(z_w-z)$ at the point $z_w$ assuming
$z$ is fixed; $\Gamma = \Gamma' + Y'_1 + Y'_2$;  coordinates $z_1$
and $z_2$ are the same as in (\ref{per_e}); and $\zeta (z_w)$ is
the external potential.
\subsubsection{Numerical
approximation} We solve the integral equation numerically, by
approximating the source densities  by step-functions
\cite{numer}. Thus we divide the given boundary $\Gamma$ into
$N_\Gamma$ intervals and assume that the simple source density
$\sigma$ has a constant value within each interval. Then denoting
these constant values by $\sigma_i$, $i=1,2,...,N_\Gamma$, we
approximate $\Phi$  and $E$ by
\begin{eqnarray}
\widehat{\Phi}(z_w) = \sum_{i=1}^{N_\Gamma}
{\sigma_i}{\epsilon_0} \int_i
\log(z_w-z) dz,~~ {\textrm{and}} \label{phi_appr}\\
\widehat{E}(z_w) = \sum_{i=1}^{N_\Gamma}
\frac{\sigma_i}{\epsilon_0} \int_i \left( \frac{d
\log(z_w-z)}{dz_w}\right)^* dz, \label{e_appr}
\end{eqnarray}
where $\int_i$ denotes integration over the $i$-th interval of
$\Gamma$. We substitute (\ref{phi_appr}) and (\ref{e_appr}) into
(\ref{per_e_int}) to obtain numerical approximation for periodic
solution. The unknowns (in the system obtained) are the charge
density on the intervals $\sigma_i$, the potential and the
electric field  on the periodic aperture. All coefficients in the
system are calculated analytically. For practical geometries, the
matrix of coefficients is well defined and there is no difficulty
in solving the system directly. For macroparticle tracking, the
electric field calculated on polar grid and  then interpolated at
the macroparticle position (same as for RF fields).
\vspace{-1mm}
\subsection{Tracking}
\vspace{-2mm}
 We find a macroparticle trajectory by using the 4th
order Runge-Kutta method for integrating the equation of motion
(\ref{eq_motion}) in polar coordinates. Then, we integrate the
complex electric field of the cavity modes along the trajectory to
find coefficients for the cavity's eigenmodes (\ref{A_s}). We
monitor energy conservation in order to verify accuracy of
calculation. For that purpose we use total energy that consists of
kinetic energy of the macroparticle and integral of static (due
to external potential and static space charge) and oscillating
(due to cavity modes and oscillating space charge) electric
fields along the trajectory. Initial charge and velocity
$\vec{v}$ are determined by a space-charge-limited-emission model
and a relaxation scheme. \vspace{-2mm}
\subsection{Algorithm}
\vspace{-2mm}
 We start simulation by
calculating dispersion the curve for the spatial period of the
device (using the {\it RF field solver}). Then, we calculate
electric fields for the eigenmodes. Next, (using the {\it Poisson
solver}) we calculate electric field due to external potential.
Next, we start iterations using {\it Tracking module} to find the
macroparticle trajectories, field integrals along the
trajectories, and electric fields due to space charge. Next, we
update the static and oscillating fields and start new iteration.
\vspace{-4mm}
\section{SUMMARY}
We have written a  C++ computer code that uses methods, described
above. Accuracy of resonant frequency calculation by {\it RF field
solver} for typical geometries is ~$\sim0.1$\%. We tested
performance of {\it Poisson solver} and {\it Tracking module} on
diode geometries (without magnetic field). We calculated diode
current with typical accuracy 2-3\% in comparison with analytical
solution. Testing of the code on cross- field devices is under
way. 

\end{document}